\newcommand{\evb}{ {\rm eV$^2$} } 

\documentclass[11pt]{article}
\usepackage{moriond,epsfig}
\def\be{\begin{equation}}
\def\ee{\end{equation}}
\def\bea{\begin{eqnarray}}
\def\eea{\end{eqnarray}}

\begin{document}
\vspace*{4cm}
\title{TELLING THREE FROM FOUR NEUTRINO SCENARIOS}

\author{ D. MELONI }

\address{I.N.F.N., Sezione di Roma I and Dip. Fisica, 
Universit\`a di Roma ``La Sapienza'', 
P.le A. Moro 2, I-00185, Rome, Italy }

\maketitle\abstracts{In this talk I will consider the possibility
of distinguish the usual three neutrino model from scenarios in which a
light sterile neutrino is also present. I will show that the confusion with
the so-called 3+1 scheme can arise in some particular region of the
parameter space whereas it is essentially absent for the 2+2 scheme.
 Then I will discuss the ambiguities
in the determination of the CP-violating phase $\delta$.}

\section{Introduction}
\subsection{Why four neutrinos?}

Experimental data from atmospheric neutrino experiments can be interpreted
in terms of neutrino oscillations if $\Delta m_{atm}^2$ is \cite{Toshito:2001dk}
of order 
$ \sim (1.6 - 4)\;10^{-3}$ \evb;
 solar neutrino experiments, instead, point towards a $\Delta m_\odot^2$ of 
order $\sim 10^{-4}\;{\rm eV}^{2}$ \cite{Ahmad:2002jz,Ahmad:2002ka}. 
The LSND data, on the other hand, would indicate 
a $\nu_\mu \to \nu_e$ oscillation with a third, very distinct, 
neutrino mass difference: $\Delta m_{LSND}^2 \sim 0.3 - 6\;{\rm eV}^2$.
Experiments such as  MiniBooNE \cite{solin}
will be able to confirm it in the near future.
If we consider the usual three neutrino family, it is
impossible to simultaneously explain the whole ensemble of the data since
there are only two independent neutrino mass differences. We should therefore
include at least one more neutrino ({\it sterile}, since it must be an
electroweak singlet to comply with the strong bound on the $Z^0$ invisible 
decay width) with a third, independent, mass difference to fit all the 
experimental data. 

In this framework, it seems quite relevant to understand 
if three-- and four--family 
models are distinguishable and, particularly, if the effect of CP violation in 
a three--family world could be mimicked by introducing one sterile neutrino 
in the mixing matrix \cite{Donini:2001xp}.

In order to answer to these questions we use a Neutrino Factory experimental
set-up \cite{Geer:1998iz,DeRujula:1999hd}, which has already been shown to be 
the best tool to explore neutrino masses and mixing (in particular, in the
four-family mixing scenario in \cite{Donini:1999jc,Donini:2001xy}). 
Without entering in the
details, in a Neutrino Factory muons are accumulated in a storage ring and then 
decay in a straight section of the experimental apparatus producing two
different flavours of neutrinos and antineutrinos, 
depending on the initial muon polarity. The advantage to follow this procedure
is the purity of the neutrino fluxes, not achievable with usual 
(super)beam. In such an experimental setup, the final muons obtained, for
example, via the chain:
\bea
\mu^+  \rightarrow e^+ & \bar \nu_\mu    & \nu_e  \nonumber \\
                       &                 &\downarrow  \nonumber \\
                       &                 &\nu_\mu \rightarrow 
		       \mu^- \nonumber
\eea
are the signal of the oscillation and are called "wrong sign muons" ($\mu^-$
appearance in a $\mu^+$ beam).

\subsection{Schemes in four neutrino scenarios }
When four neutrinos are considered, two very different classes of mass
spectrum are possible: three almost degenerate neutrinos, accounting for
the solar and atmospheric oscillations, separated from the fourth one by the
large LSND mass difference (3+1 scheme); or two almost degenerate neutrino
pairs, accounting respectively for the solar and atmospheric oscillations,
separated by the LSND mass gap (2+2 scheme).
The recent analysis of the LSND experimental data reconciles the 3+1 scheme
with exclusion bounds coming from other reactor and accelerator experiments,
so that the 3+1 model is now marginally compatible with the 
data. Moreover the recent SNO results restricted the allowed parameter region 
for the 2+2 scheme giving a considerably
worse fit to the experimental data with respect to the pre-SNO analysis 
\cite{Foglid64,Gonz}. \\ 
In the following we will focus our
attention only on the 3+1 scheme, which has the three neutrino model as a
limit in the case of vanishing gap-crossing angles and therefore it can more
easily be confused with it.\\ 

\section{Three or four families?}
In a few years from now the LSND results will be confirmed by MiniBooNE 
or not. In case of a non-conclusive result, 
the three-family mixing model will be considered the most plausible
extension of the Standard Model, so that long baseline experiments
will be preferred with respect to the (four-family inspired) short baseline 
ones. In this case, will a Neutrino Factory and corresponding detectors, 
designed to explore the three-family mixing model, be able to tell three 
neutrinos from four neutrinos?

\subsection{Experimental set-up and strategy adopted}
We consider the following ``reference set-up'': neutrino beams resulting from 
the decay of $ 2 \times 10^{20} \mu^+$'s and $\mu^-$'s per year in a straight 
section 
of an $E_\mu = 50$ GeV muon accumulator. A realistic 40 Kton 
detector of magnetized iron is used and five years of data taking 
for each polarity is envisaged. Detailed estimates of the corresponding 
expected backgrounds ($b$) and efficiencies ($\epsilon$) have been included in 
the analysis \cite{Cervera:2000vy}. 
We follow the analysis in energy bins as made in 
\cite{Cervera:2000kp,Donini:2000ky,Burguet-Castell:2001ez}. 
The $\nu_e \to \nu_\mu$ channel, the so--called {\sl golden channel}, 
will be the main subject of our investigations.

Let $N^{i}_{4 \nu}$ be the total number of wrong-sign muons in a four neutrino
theory detected 
when the factory runs in polarity  $\mu^+$ or $\mu^-$, grouped in energy bins 
specified by the index $i$, and three possible 
distances, corresponding to $L = 732$ Km, $L = 3500$ Km and 
$L = 7332$ Km.
In order to simulate a typical experimental situation we generate 
a set of ``data'' $n^i$ by smearing the number of wrong sign muons:

\begin{eqnarray}
n^i = \frac{ {\rm Smear} (N^i_{4 \nu} \epsilon^i + 
b^i) - b^i}{\epsilon^i} \,. 
\end{eqnarray}

Finally, the "data" are fitted to the theoretical expectation in the
three-neutrino model as a function of the neutrino parameters under study, 
using a $\chi^2$ minimization:
\be
\chi^2 = \sum_i 
\left(\frac{ n^i \, - \, 
N^i_{3 \nu}}{\delta n^i}\right)^2 \, ,
\label{chi2}
\ee
where $\delta n^i$ is the statistical error for 
$n^i$ (errors on background and efficiencies are neglected).
The output of interest of this procedure are the values of the $\chi^2 \le
\chi^2_{68 \%,\, n\, dof}$, for which the hypotesis of confusion can be 
considered accepted.\\
In order to get results that can be easily understood, we
need to restrict the 4-$\nu's$ parameter space in our numerical simulations: 
the $4 \times 4$ unitary mixing matrix has six independent angles and 
three CP-violating phases! We only considered
CP-conserving four neutrino schemes. We fixed $\theta_{12}=\theta_\odot=
22.5^\circ$ and $\theta_{23}=\theta_{atm}= 45^\circ$. Then we allowed a 
variation for $\theta_{14}=\theta_{24}=\epsilon=2^\circ,~5^\circ$ and $10^\circ$ 
(taken to be equal for simplicity) and the two free parameters  
we considered are $\theta_{13}$ in the interval 
$\left[1^\circ,\, 10^\circ  \right]$ and $\theta_{34}$ in 
$\left[0^\circ,\, 50^\circ  \right]$ \cite{Bahcall:2002zh}. 
For the three family case, we still fixed 
$\theta_{12}=\theta_\odot$ and $\theta_{23}=\theta_{atm}$ since the differences
 from the two models depend on the small $\Delta_\odot$ and on 
the angles $\theta_{14}$, $\theta_{24}$ and $\theta_{13}$ that we took small.
In this case the parameter space consists of the angle $\theta_{13}$ 
(varying in the same
interval as in the 3+1 scheme) and the CP-violating phase $\delta$ in the interval 
$\left[-180^\circ,\, 180^\circ  \right]$.

\begin{figure}[h!]
\begin{center}
\epsfxsize15cm\epsffile{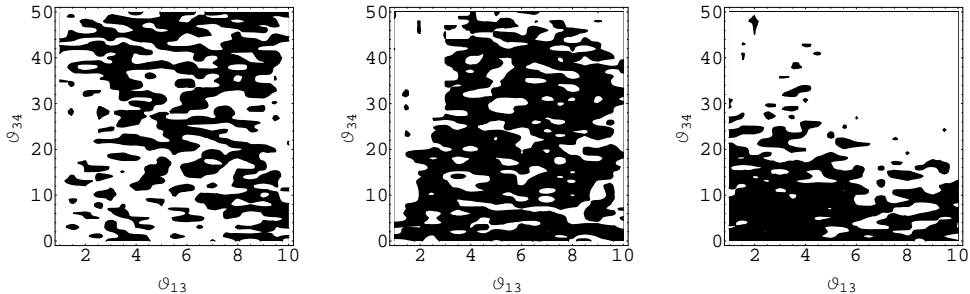}
\caption{ Plots at 68 \% in the four--family plane for different
 baselines for $\theta_{14}=\theta_{24}=2^{\circ}$.
From left to right:\hspace{1truecm}
 L= 732 Km; L = 3500 Km; L = 7332 Km. 
\label{fig:dalm=2}}
\end{center}
\end{figure}

The results of the fits depend heavily on the values of the small
gap-crossing angles $\theta_{14}$ and $\theta_{24}$. If their value is small 
($2^{\circ}$), since the 3+1 model has a smooth limit to the three--neutrino 
theory, the fit is possible for almost every value of the other parameters.
This is shown in fig.~\ref{fig:dalm=2}, where in the dark regions of the 
``dalmatian dog hair'' plot the three--neutrino model is 
able to fit at 68\%  c.l. the data generated with those parameter values. 

\begin{figure}[h!]
\begin{center}
\epsfxsize15cm\epsffile{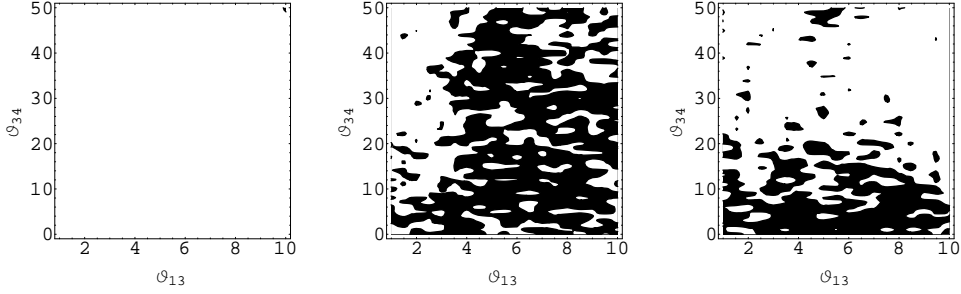}
\caption{ Plots at 68 \% in the four--family plane for different
 baselines for $\theta_{14}=\theta_{24}=5^{\circ}$.
From left to right:\hspace{1truecm}
 L= 732 Km; L = 3500 Km; L = 7332 Km.
\label{fig:dalm=5}}
\end{center}
\end{figure}

Increasing the value of $\theta_{14}$ and $\theta_{24}$, 
the extension of blotted regions decreases. 
This is shown for $\theta_{14}=\theta_{24}= 5^{\circ}$ in
fig.~\ref{fig:dalm=5}.

A further increase of the gap-crossing angles makes the distinction of the
two models possible for all values of the other, variable parameters.

Assuming a detector with better resolution, (i.e. increasing the number of 
energy bins from five to ten) 
we have not observed a sensible reduction of the blotted regions 
for $ \epsilon = 2^\circ$, due to the extremely poor statistics per energy
bin. On the other hand, for $ \epsilon = 5^\circ$ we observe a sensitive 
reduction of the confusion regions.

To explain all the previous results, we can consider the following plots in
which we represented the transition probabilities for the two models for the 
oscillation parameters fixed to some representative value 
(fig. \ref{fig:explanation}):

\begin{figure}[h!]
\begin{center}
\begin{tabular}{cc}
\epsfxsize7cm\epsffile{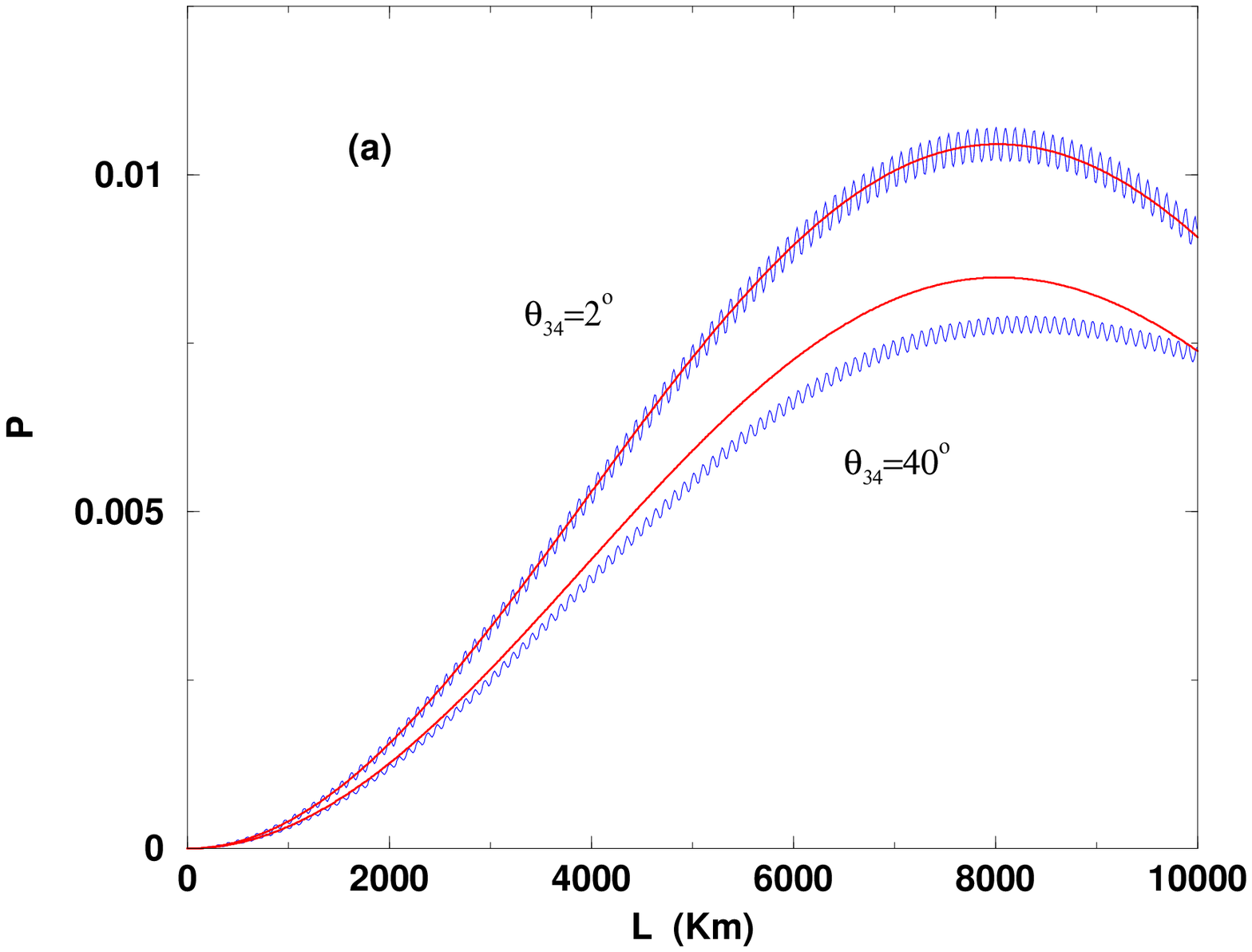} &
\epsfxsize7cm\epsffile{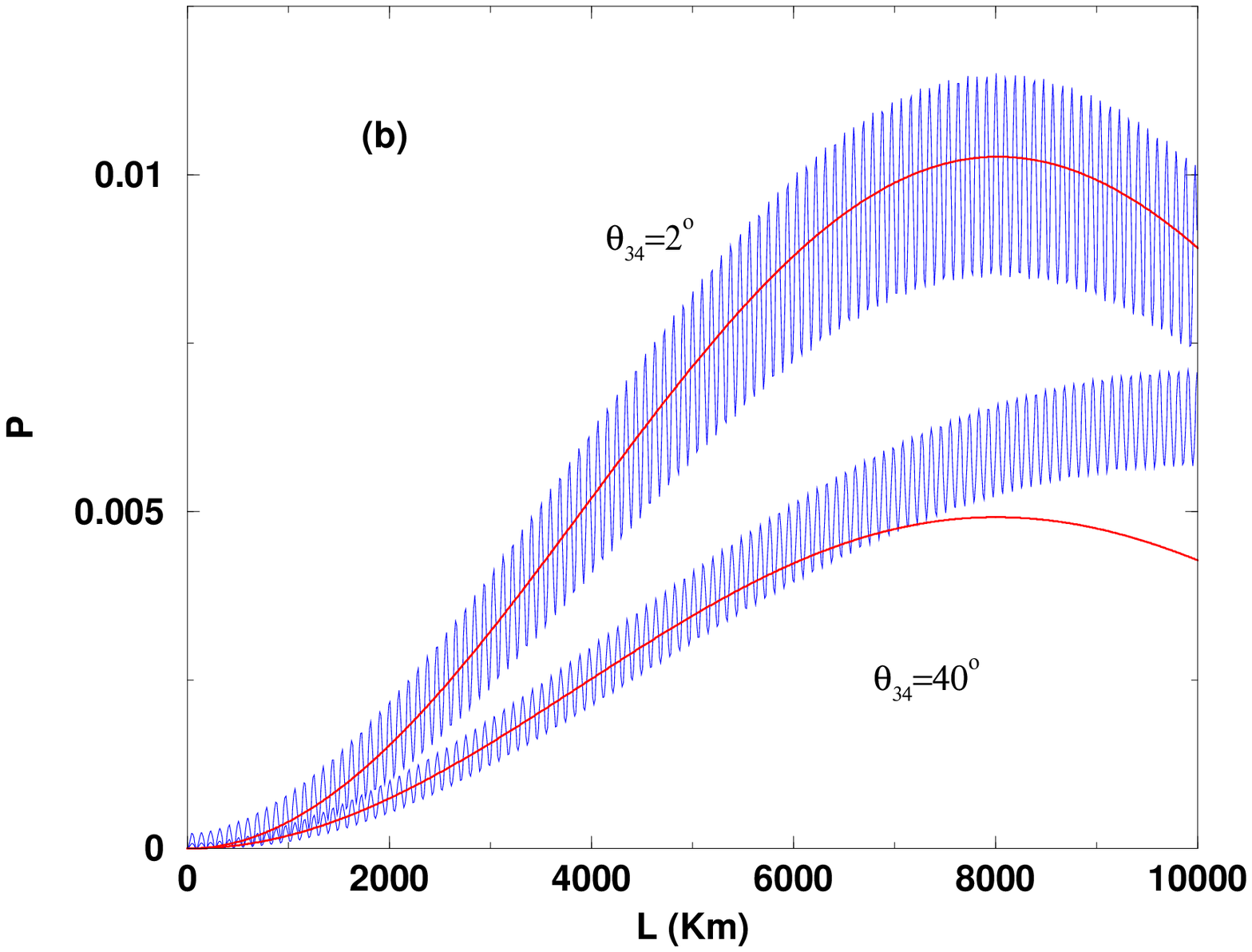} 
\end{tabular}
\caption{ Oscillation probabilities for $\nu_{e} \to \nu_{\mu}$ as a 
function of the pathlength L for $E_\nu$ = 38 GeV and (a) $\epsilon 
=2^{\circ}$, (b) $\epsilon =5^{\circ}$. The solid lines refer to the
three-neutrino model whereas the wiggled ones refer to 3+1 scheme.
In this case we put $\theta_{13}=8^\circ$ 
 and $\delta=0^\circ$.
\label{fig:explanation}}
\end{center}
\end{figure}

For $\epsilon =2^{\circ}$ (plot on the left) the four--neutrino 
probability is
very similar to the three--neutrino result and confusion is always possible.
At low and intermediate distances ($L \sim $ 3000 Km) the oscillation 
probabilities in the two models are quite similar: at this distance we expect 
therefore that it will be difficult to tell three from four neutrinos. 
At larger distances ( $L > $ 5000 Km) the distinction will in general 
be possible for $\theta_{34}$ large enough.

For $\epsilon =5^{\circ}$ (plot on the right), at the shortest distance, we can
observe that the LSND oscillation dominates in the oscillation probability: 
the 3+1 model gives a probability larger than 
the three--neutrino theory and confusion is not possible. 
The oscillation probabilities at the 
intermediate distance are very similar in the two models, and the 
confusion is therefore maximal in this case. At larger distances the situation
remains essentially unchanged.

\section{CP violation vs. more neutrinos}

\begin{figure}[h!]
\begin{center}
\begin{tabular}{ccc}
\hspace{-1cm} \epsfxsize5.2cm\epsffile{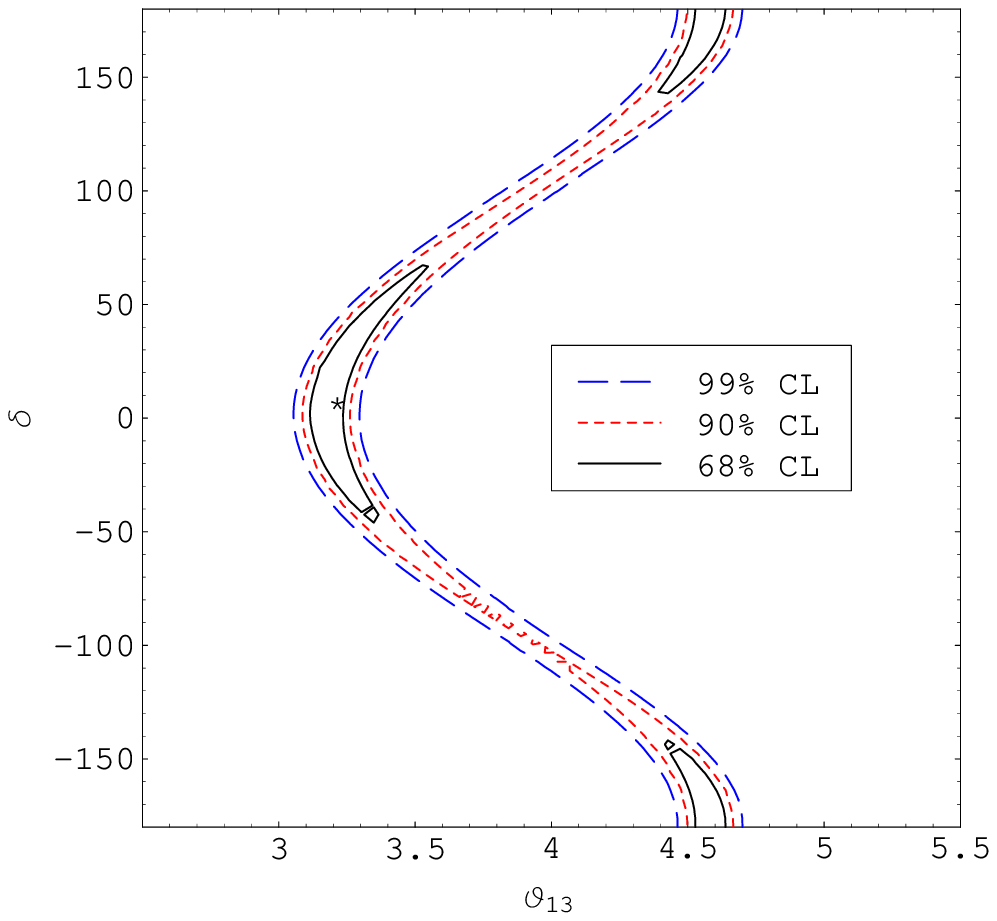} &
\epsfxsize5.2cm\epsffile{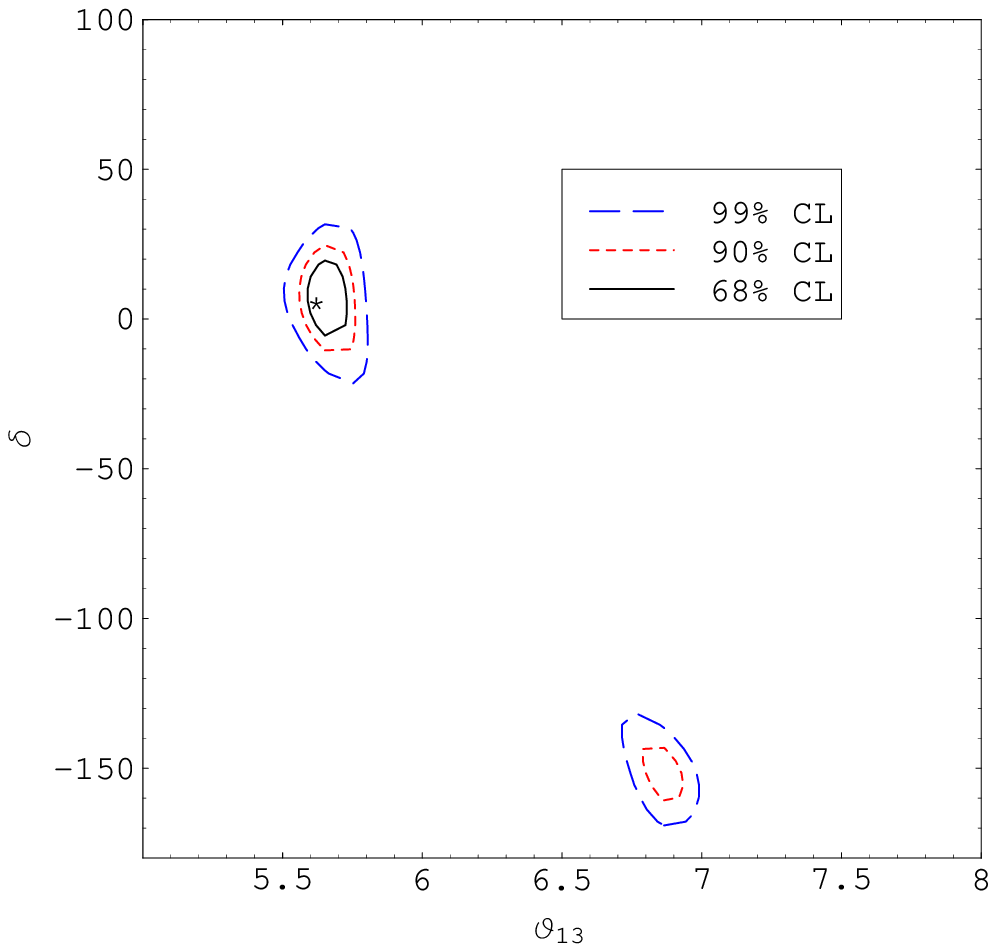} &
\epsfxsize5.2cm\epsffile{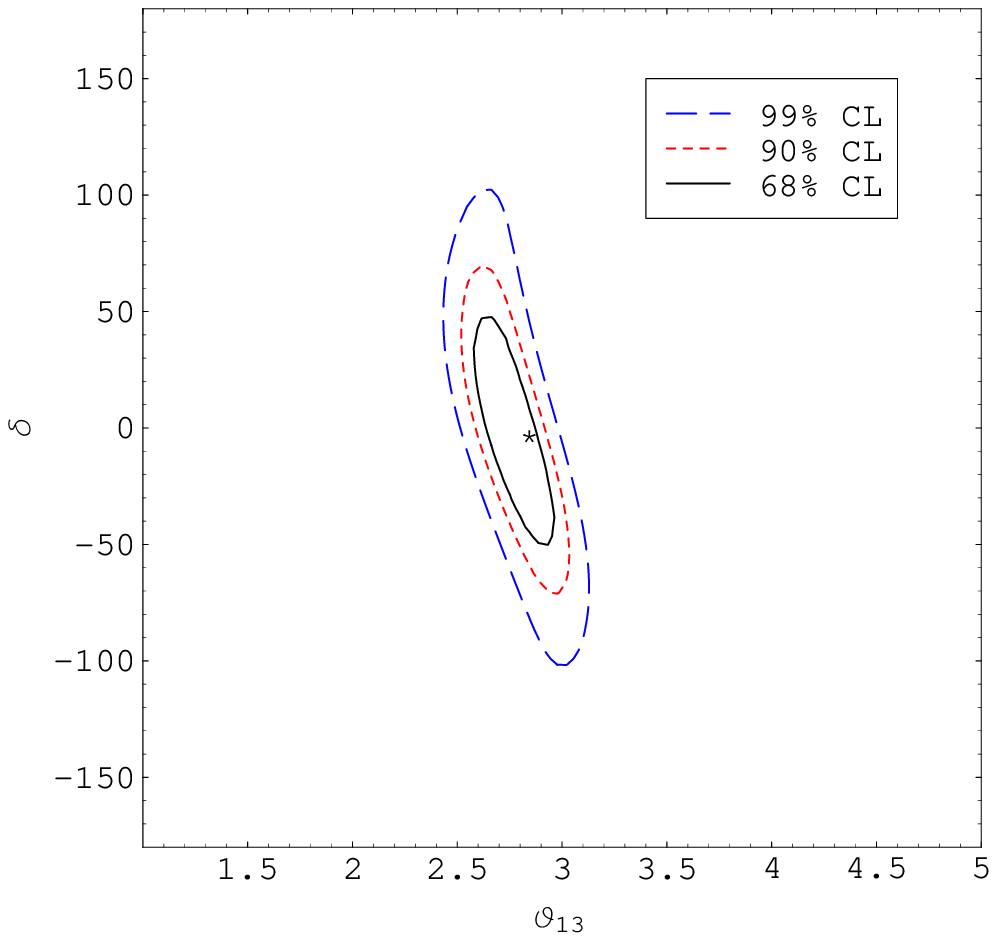}
\end{tabular}
\caption{\label{fig:3-CP-fit}
Confidence level contours for typical points where the
three--neutrino theory can well reproduce (3+1)--neutrino ``data'' at the
three distances studied: from left to right, $L=732$, 3500 and 7332 Km.}
\end{center}
\end{figure}

So far, we have not
presented in detail the results of a fit of the four--family ``data'' 
in the parameter space of the three-family model.
If a non-vanishing CP violating phase is found when fitting with the 
three--family model the 3+1 (CP conserving) ``data'', its values are generally 
not too large. To illustrate this conclusion, we show the situation in the three
family parameter space for a tipical point in the dark region
(fig. \ref{fig:3-CP-fit}), where we present the 68\%, 90\% and 99\% confidence level 
contours for typical three--neutrino fits (one for each distance) 
to 3+1 neutrino ``data'' 
generated with $\theta_{14}=\theta_{24}=2^\circ$ and five energy bins. 

The representativeness of the values chosen 
for the plots in fig.~\ref{fig:3-CP-fit} can be understood noting  
that, out of about 6000 successful fits, 31\% (50\%, 37\%) for 
$L =732~(3500, 7332)$ Km give for the CP violating phase a value 
$-15^\circ < \delta < 15^\circ$, that is the amount of the leptonic
CP-violation is not large.

\begin{figure}[h!]
\begin{center}
\begin{tabular}{ccc}
\hspace{-1cm}
\epsfxsize5cm\epsffile{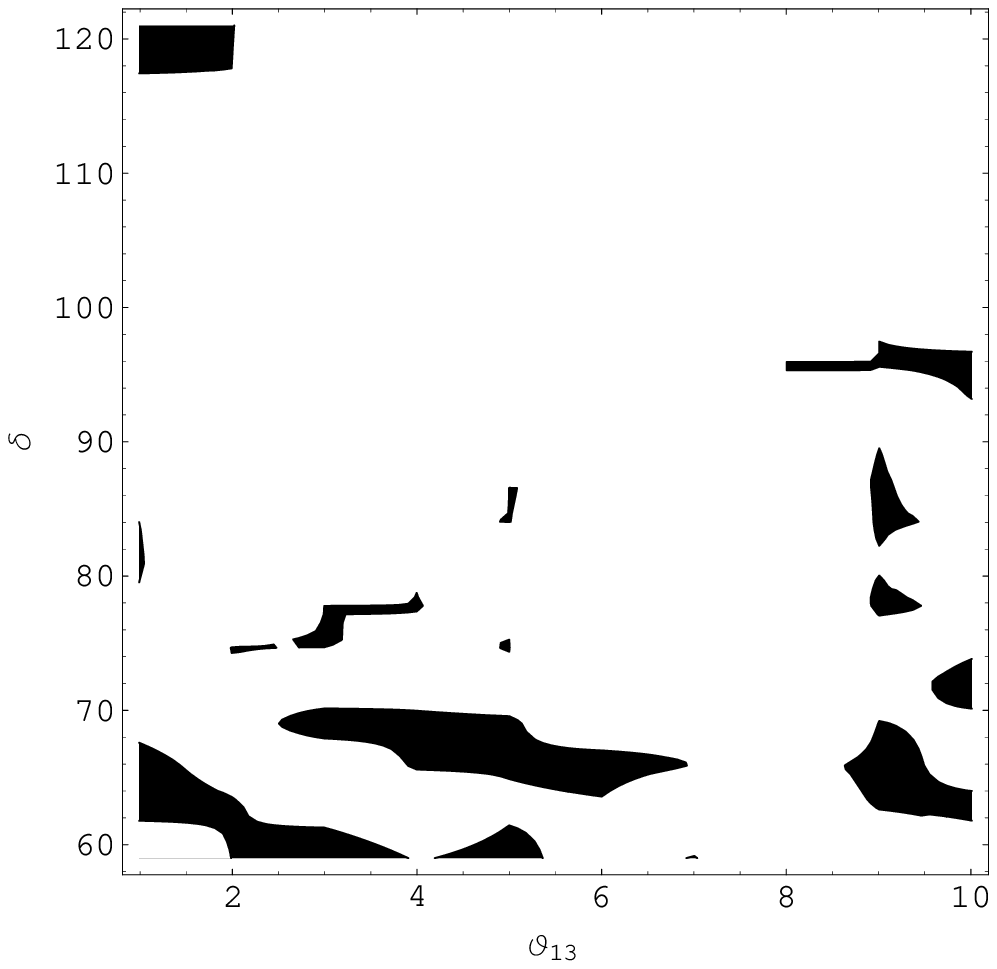} &
\epsfxsize5cm\epsffile{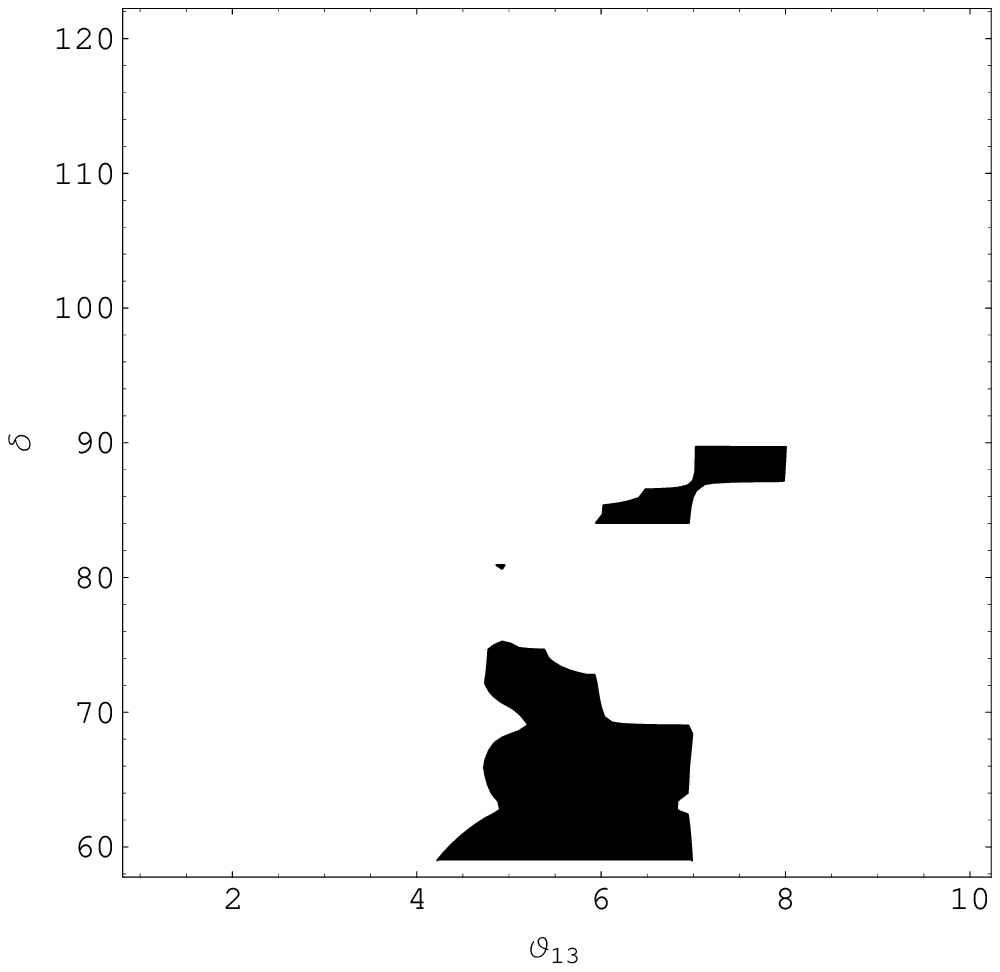} &
\epsfxsize5cm\epsffile{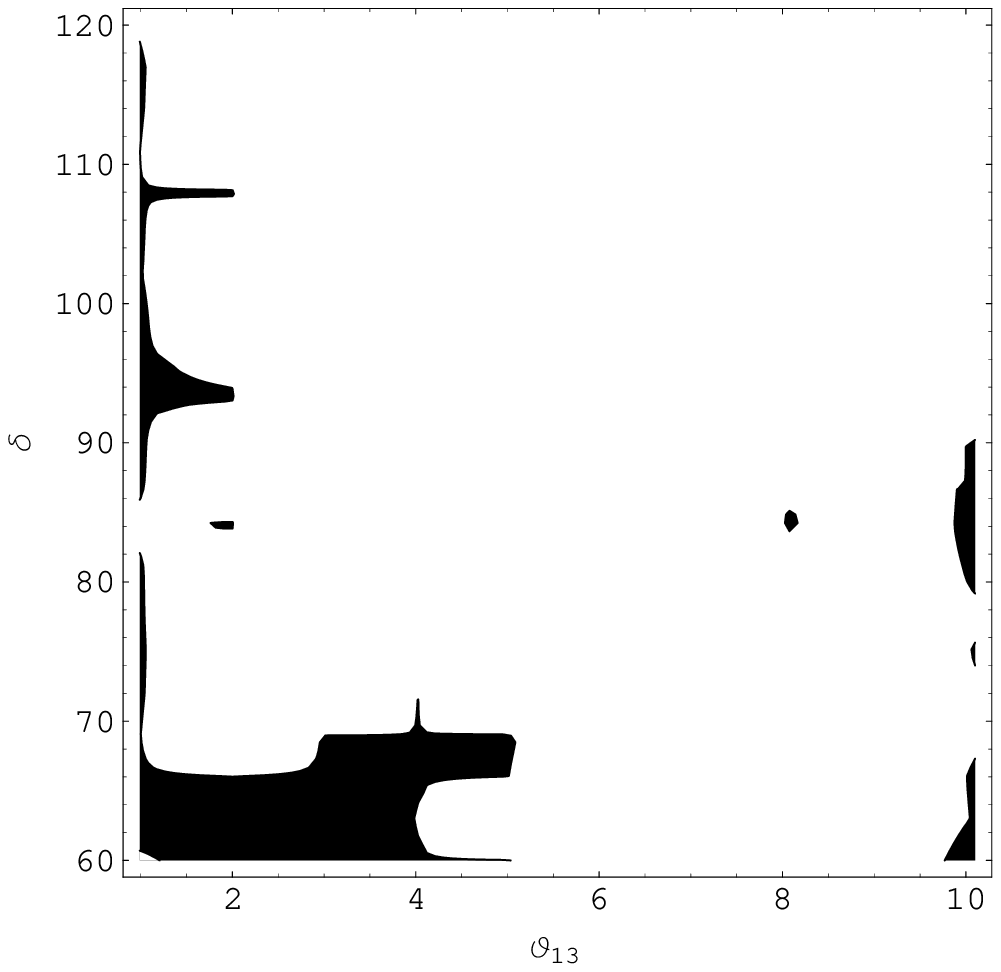}
\end{tabular}
\end{center}
\caption{Examples of regions in the ($\theta_{13}$, $\delta$) plane  
where the (3+1)--neutrino theory with vanishing CP violating phases can 
reproduce at 90\% c.l. three--neutrino ``data''.
From left to right: L = 732; L = 3500; L=7332.
\label{fig:3in4}}
\end{figure}

Suppose now to follow an inverse procedure with respect to that described 
previously, that is to generate "data" in the usual three-neutrino model; 
can we miss a large CP-violating phase in the 
three-family model by fitting in a CP-conserving 3+1 theory? For this purpose we 
choose to vary $1^\circ \leq \theta_{13} \leq 10^\circ$ and
$60^\circ \leq \delta \leq 120^\circ$.
The answer to the question depends on the baseline, 
as we can see in fig. \ref{fig:3in4}.

The confusion is 
possible in many a case for $L$ = 732 Km, it is very difficult to 
obtain for the intermediate distance, $L$ = 3500 Km, and at the largest $L$
an intermediate situation holds. Fitting simultaneously data at two different 
distances the possibility of confusion is strongly reduced and in fact 
it vanishes if the data at intermediate distance are used in any 
combination with the others. We also observed that no confusion would be 
possible assuming the larger values $\epsilon$ = 5$^\circ$ or 
10$^\circ$. 

Our result seems to imply that if the data would point to a maximal 
CP violating phase in the three--neutrino theory, it would be very 
difficult to describe them in a theory without CP violation, 
even if with more neutrinos.  

Eventually in the 2+2 scheme the ambiguity with a three neutrino theory
is essentially absent due to the impossibility to recover the three family
formulae in any limit.

\section*{Acknowledgments}
I am particularly indebted with prof. M. Lusignoli and A. Donini for the
useful discussion and the Organizer Committee for the stimulating atmosphere of
the Moriond Conference.

\end{document}